\documentclass[fleqn,usenatbib]{mnras}

\usepackage{newtxtext,newtxmath}

\usepackage[T1]{fontenc}

\DeclareRobustCommand{\VAN}[3]{#2}
\let\VANthebibliography\thebibliography
\def\thebibliography{\DeclareRobustCommand{\VAN}[3]{##3}\VANthebibliography}

\usepackage{graphicx}	
\usepackage{amsmath}	

\title[Tides in clouds]{Tides in clouds: control of  star formation  by long-range gravitational force}

\author[Guang-Xing Li]{
Guang-Xing Li $^{1}$\thanks{E-mail: ligx.ngc7293@gmail.com, gxli@ynu.edu.cn} \\
$^{1}$ South-Western Institute for Astronomy Research, Yunnan University,
Kunming, 650500 Yunnan, People's Republic of China}

\date{Accepted XXX. Received YYY; in original form ZZZ}

\pubyear{2015}

\begin{document}
\label{firstpage}
\pagerange{\pageref{firstpage}--\pageref{lastpage}}
\maketitle

\begin{abstract}
	Gravity drives the collapse of molecular clouds through which stars form, yet the exact role of gravity in
	cloud collapse remains a complex issue.  Studies point to a picture where
	star formation occurs in clusters.
	In a typical, pc-sized cluster-forming region, the collapse is hierarchical, and the stars should be born from regions of 
	even smaller sizes ($\approx 0.1\;\rm pc$). The origin of
	this spatial arrangement remains under investigation. Based on a high-quality surface density map towards the Perseus region, we construct a 3D density
	structure, compute the gravitational potential, and derive eigenvalues of the tidal tensor
	($\lambda_{\rm min}$, $\lambda_{\rm mid}$, $\lambda_{\rm max}$, $\lambda_{\rm
	min} < \lambda_{\rm mid} < \lambda_{\rm max}$),  analyze the behavior
	of gravity at every location and reveal its multiple roles in cloud evolution. We find that fragmentation is limited to several isolated, high-density ``islands''.  Surrounding them, is a vast amount of the gas ($75 \%
	$ of the mass, $95 \%$ of the volume) stays under the influence of extensive
	tides where fragmentation is suppressed.  
This gas will be transported towards these regions to fuel star formation.  
	The spatial
	arrangement of regions under different tides explains the hierarchical and localized pattern of star
	formation inferred from the observations.  Tides were first recognized by Newton, yet this is the first time its dominance in cloud evolution has been revealed. 
	We expect this link between cloud density structure and role gravity to be strengthened by future studies, resulting in a clear view of the star formation process.
\end{abstract}


\section{Introdcuction}
\begin{figure*}
  \includegraphics[width= 0.8 \textwidth]{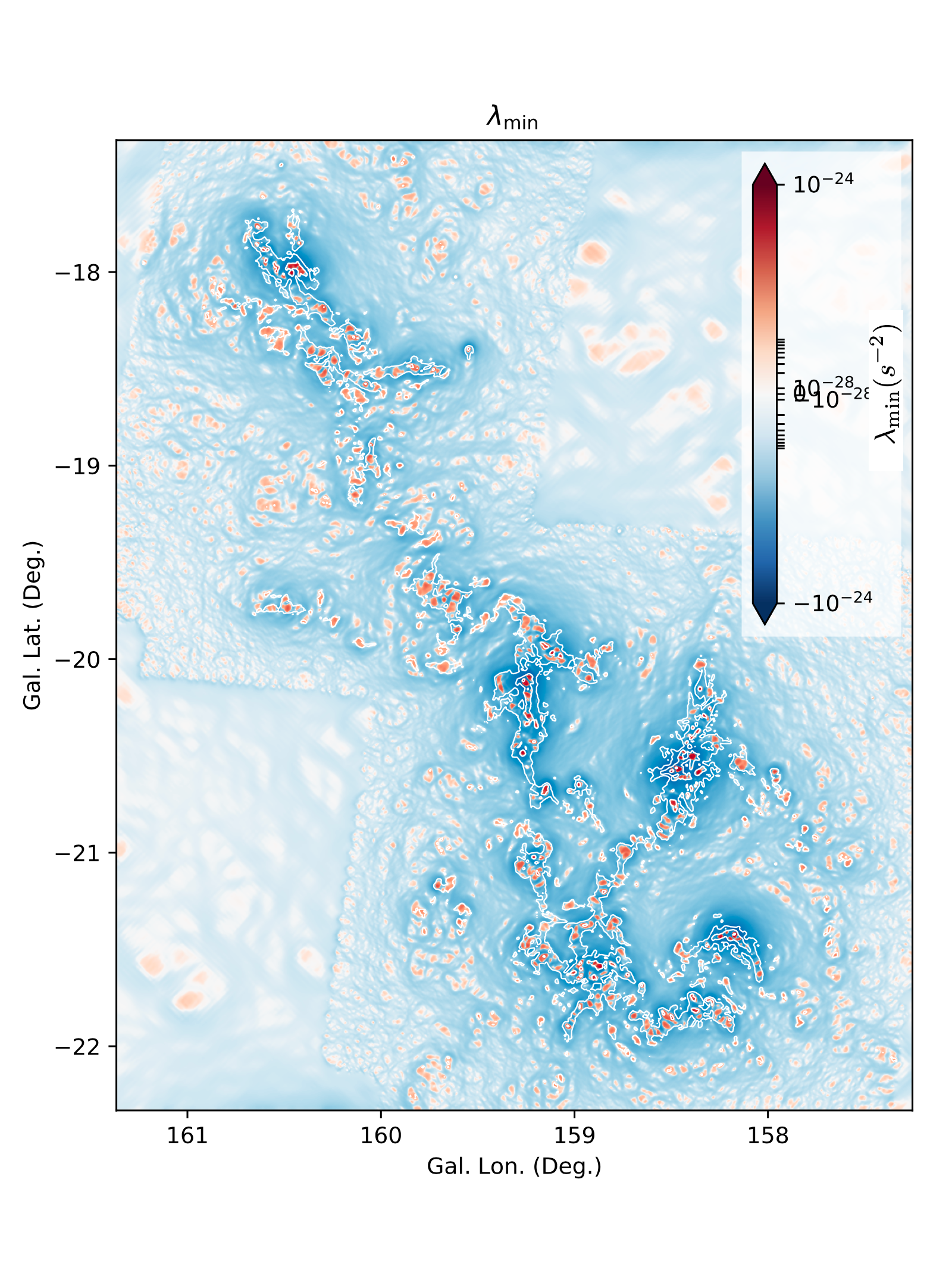}
\caption{\label{fig:eigs1} {\bf Maps of the eigenvalues of the tidal tensor $\lambda_{\rm min}$  towards the Perseus molecular cloud. Seen from the map of $\lambda_{\rm min}$, for the majority of the volume, the tides are extensive and fragmentation is suppressed.}  Thin while lines: $n_{\rm H_2} = 3.1 \times 10^{3, 4, 5}\;\rm cm^{-3}$. Thick lines in the $\lambda_{\rm max}$ map: Boundaries of some subregions. The plots are taken at the central plane of a 3D density distribution reconstructed from observations (Appendix \ref{sec:reconstruction}). }
\end{figure*}

\begin{figure*}
	\includegraphics[width= 0.8 \textwidth]{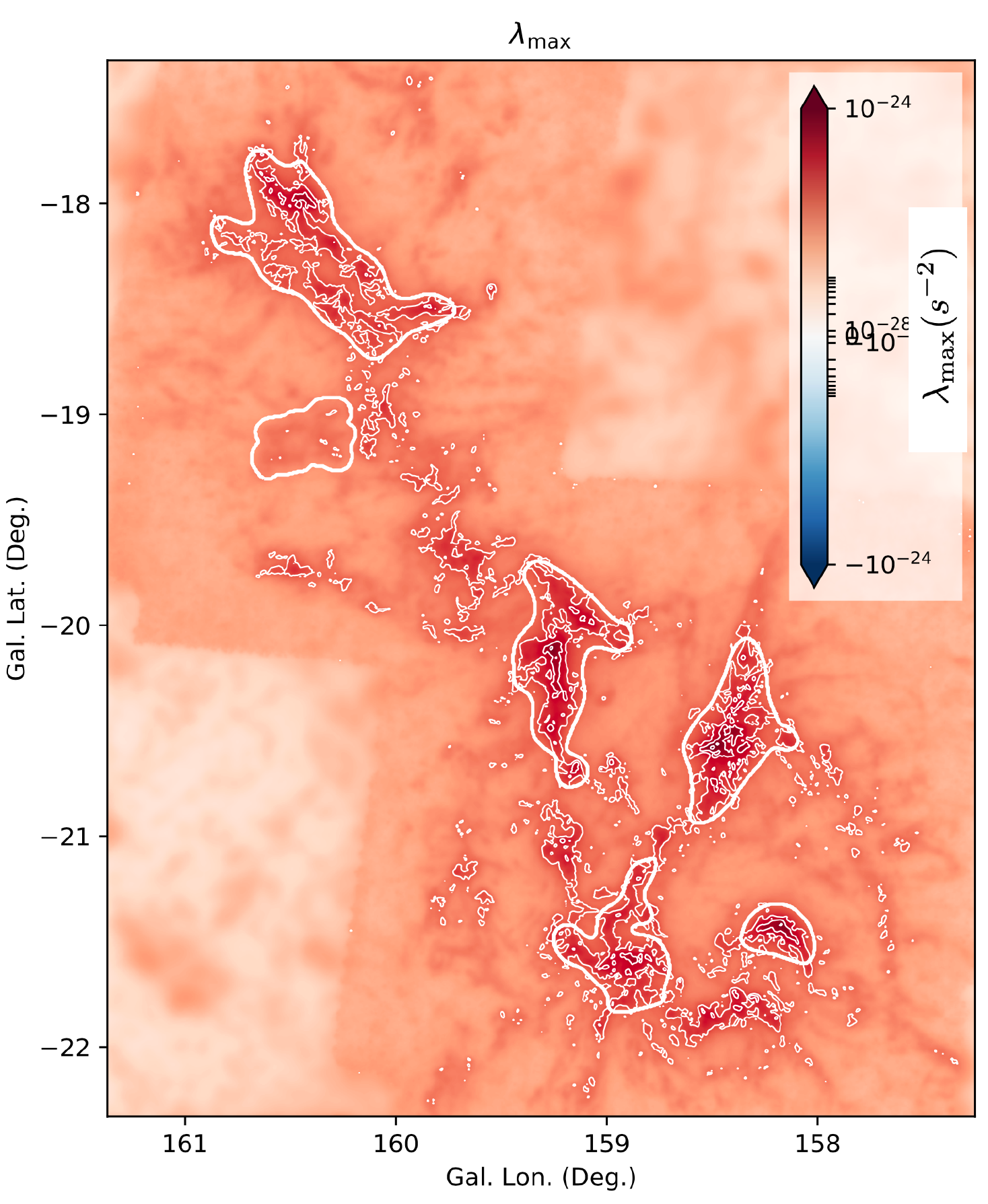}
  \caption{\label{fig:eigs3} {\bf Maps of the eigenvalues of the tidal tensor $\lambda_{\rm max}$ towards the Perseus molecular cloud.}  Thin while lines: $n_{\rm H_2} = 3.1 \times 10^{3, 4, 5}\;\rm cm^{-3}$. Thick lines in the $\lambda_{\rm max}$ map: Boundaries of some subregions. The plots are taken at the central plane of a 3D density distribution reconstructed from observations (Appendix \ref{sec:reconstruction}). }
  \end{figure*}



{M}olecular clouds are
complex objects that exhibit significant fluctuations on all scales
\citep{2000prpl.conf...97W}. These cloud surface densities often exhibit variations of multiple orders of magnitudes, and are often quantified using log-normal or
power-law models \citep{2009A&A...508L..35K,2014Sci...344..183K}.  The collapse of the molecular cloud is a 
complex, multi-scale process that involves an interplay between turbulence, gravity, magnetic field, and ionization \citep{2014prpl.conf....3D}. 
The newly-formed stars are also organized in hierarchies: Up to 85\% of  
the young stars should
form in clusters\citep{2016AJ....151....5M}. Stars in one cluster 
are likely to be assembled from stars formed in
groups \citep{2017MNRAS.467.1313V}  which contain small numbers of stars. 
By modeling the distribution of orbital
parameters, a study \citep{2012A&A...543A...8M} find that stars in
clusters are born in regions of surprisingly high stellar mass densities ($\rho = 3\times 10^3
M_{\odot}\;\rm pc^{-3}$ for a 300 $M_{\odot}$ cluster). What controls the evolution of these density structures through which stars form remains an open question.  


Gravity can play multiple roles in cloud evolution: Inside dense regions, it
drives fragmentation and collapse, and outside these regions, it can suppress
the low-density gas from collapsing through extensive tidal forces and drive
accretion. Past studies focus on the role of gravity in individual, isolated parts,
neglecting long-range gravitational interactions.  This negligence is partly caused by the picture people have in mind:
In the prevailing paradigm for star formation,  gravity drives the collapse, balanced by supports from turbulence and magnetic
fields \citep{2019ARA&A..57..227K,2020SSRv..216...64K}. This picture focuses people's attention on the interplay between e.g. turbulence
and gravity,  neglecting the fact that role of gravity can be different
depending on the condition. Another reason is methodological, as the most
widely-adopted way to quantify the importance of gravity is to evaluate the
virial parameter \citep{1992ApJ...395..140B}, which is the ratio between kinetic
and gravitational energy, where gravitational interactions between matter from the inside and the
outside of these boundaries are neglected.


The tidal tensor, defined as $T_{ij} = \frac{\partial \phi}{\partial i \partial
j}$, where $\phi(x, y, z)$ is the gravitational potential, provides a detailed
description to the local structure of a gravitational potential. The eigenvalues
of the tidal tensor contain information on how  a region would evolve
under gravity.
 Under self-gravity, $\lambda_{\rm min} \approx \lambda_{\rm max} \approx
\lambda_{\rm mean} \approx 4/3 \pi G \rho$, where a Jeans-like fragmentation
should occur \citep{1902RSPTA.199....1J}. Adding gravity from external bodies,
$\lambda_{\rm min}$ becomes smaller. When $0\lesssim \lambda_{\rm
min}<<\lambda_{\rm mean}$, the fragmentation is slowed down, accompanied  by an
increased Jeans mass \citep{2013MNRAS.434L..56J}, and when $ \lambda_{\rm min}
<0$, Jeans-like fragmentation becomes impossible. Although the cloud collapse is
also under the influences from e.g. turbulence \citep{2004RvMP...76..125M} and
magnetic field \citep{Li2014},  their role is to provide support against gravity and is secondary compared to that from gravity.
Thus, a solid understanding of the behavior of gravity should provide direct
insights on how collapses occur and help to establish the connection between the
density structure of the cloud and star formation.  
 
Deriving the tidal tensor for observed clouds has been a challenging  task, as
this would require 3D density distributions, whereas current observations only
provide 2D information in high resolutions. Taking advantage of a density
reconstruction method developed from our previous papers, we perform a first
study of the structure of the gravitational field towards a real molecular
cloud.

We use the surface density map constructed by combining Herschel and Planck observations \citep{2016A&A...587A.106Z}.   The map has a spatial resolution of 0.03 pc.  We construct a 3D density distribution (Appendix \ref{sec:reconstruction}), which shares many features with the original density structure.  Using results from a 
numerical simulation performed by \citep{2019MNRAS.486.4622C}, we verify that these differences are small (Appendix \ref{sec:comparison}), and this reconstruction  allows us to
perform the analysis presented below.
\begin{figure*}
  \includegraphics[width= 1\textwidth]{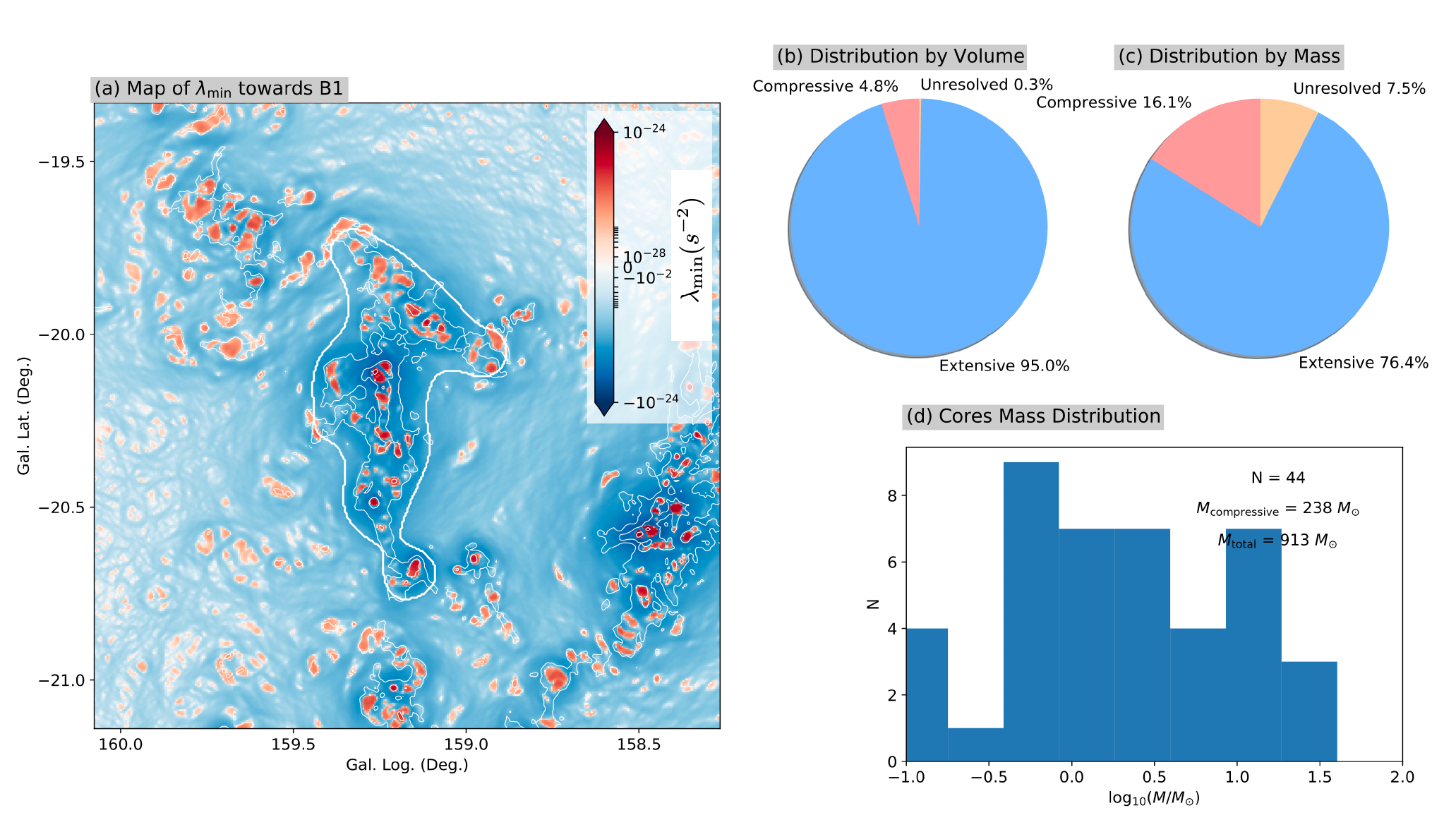}
\caption{  {\bf  Importance of extensive tides in the B1 region in the Perseus molecular cloud.} (a) A map of $\lambda_{\rm min}$ towards the B1 region. Blue stand for extensive tides ($\lambda_{\rm min}<0$) (b) a breakdown of the volume by the properties of tides. (c) a breakdown of mass. (d) mass spectrum of coherent regions under compressive tides. 
 \label{fig:b1:extensive}}
\end{figure*}

\section{Results}
We compute gravitational potential by solving the Poisson equation 
\begin{equation}
	\nabla^2 \phi = 4 \pi G \rho \;,
\end{equation}	
	 and derive the tidal tensor  (Appendix \ref{sec:phi}) 
\begin{equation}
	T_{ij} = \frac{\partial^2 \phi} { \partial i \partial j} \;,
\end{equation}
The tidal tensor is a 3x3 symmetric matrix, which have three eigenvalues $\lambda_{\rm min}$,  $\lambda_{\rm
mid}$, $\lambda_{\rm max}$, where $\lambda_ {\rm min}<\lambda_{\rm mid} <
\lambda_{\rm max}$. The three eigenvalues controls the gas evolution along three orthogonal directions. The tide is \emph{fully compressive} if 
\begin{equation}
\lambda_{\rm min} >0,\; \lambda_{\rm mid} > 0, \; \lambda_{\rm max} >0 \;,
\end{equation}
and is \emph{extensive} if one of the eigenvalues is smaller than zero, e.g. 
\begin{equation}
	\lambda_{\rm min} <0
\end{equation}
 The maps are plotted in Figs. \ref{fig:eigs1} \ref{fig:eigs3}.  


Since our observations have a limited resolution,  small-scale density
fluctuations are unresolved. This limitation, which is hard to overcome, can be quantified.
We adopted the criterion that gas in a voxel is resolved if the linear resolution is larger than the Jeans length 
\citep{1902RSPTA.199....1J}
($l > l_{\rm Jeans}$), and analyze the unresolved part separately.  In our case, the map has an angular resolution of 36'', and a linear resolution of $l = 0.05 \rm\;
pc$. A voxel is expected to contain significant amount of unresolved structure
if the resolution larger than the Jeans mass, 
\begin{equation}
   {\displaystyle l >  \lambda_{\text{J}}={\frac {c_{s}}{(G\rho )^{\frac {1}{2}}}}\approx 0.4{\mbox{ pc}}\cdot {\frac {c_{s}}{0.2{\mbox{ km s}}^{-1}}}\cdot \left({\frac {n}{10^{3}{\mbox{ cm}}^{-3}}}\right)^{-{\frac {1}{2}}}} \;,
\end{equation}
and gas with $n_{\rm H_2} > 10^5\;\rm cm^{-3}$ are unresolved, where a temperature of 20 K was assume. 
 When the resolution is limited the amount 
of density fluctuations is under-estimated.  In most cases, this leads to over-estimations of the amount of gas contained in regions under compressive tides. 

 


\subsection{Suppressing fragmentation by extensive tides}

When the tides are extensive, fragmentation can be suppressed. This effect has been overlooked in the past, but can be conveniently studied using our approach. The extensive tides are dominant in the regions where  $\lambda_{\rm min} < 0$ (Fig. \ref{fig:b1:extensive}). 
 To make our values well-defined, we focus on regions where $\rho_{\rm H_2} > 1000 \;\rm
cm^{-3}$ which contains to  40\% of the gas in the Perseus region. We further
divide the cloud into a few pc-sized ``clumps'' and study the composition of
gas within.  In a typical clump like the B1 (Fig. \ref{fig:b1:extensive}),  75\% (Method \ref{sec:comparison}) of the mass is
under extensive tides, and this gas occupy 95\% of the volume. These mass fractions are accurate to around 10\% (Appendix \ref{sec:comparison}). 
 To our surprise,  Long-range gravity can influence the evolution of a large majority of the gas.

\begin{figure*}
  \includegraphics[width= 1\textwidth]{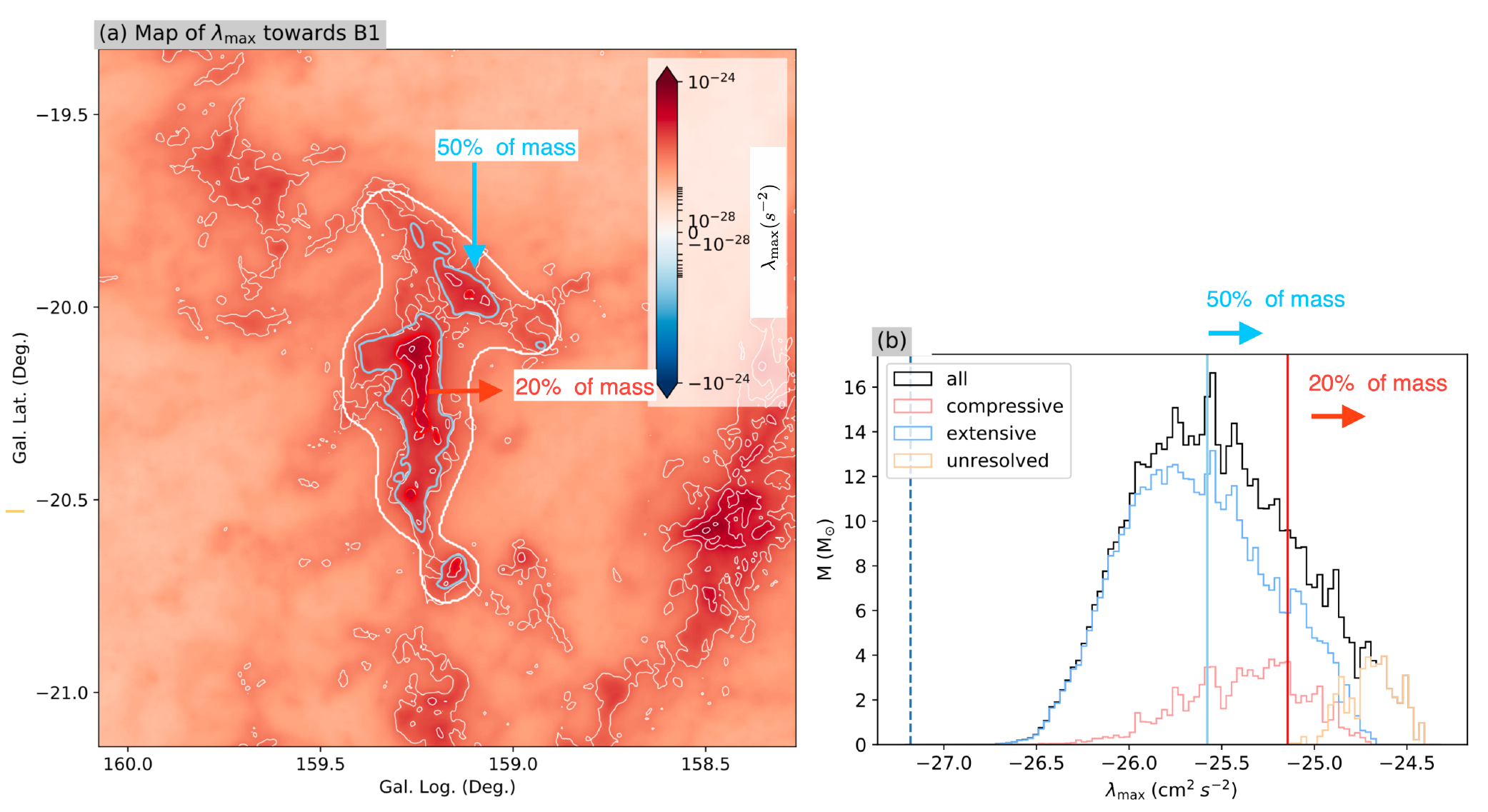}
\caption{ {\bf Maps and distributions of $\lambda_{\rm max}$.} (a) maps of
$\lambda_{\rm max}$ towards the B1 region. Thick white contour: boundary of the
region.  Red contour: boundary of the infalling regions which contains 20\% of the mass; blue contour: boundary of the infalling region which contains 50 \% of the mass. 
(b)
mass-weighted distribution of $\lambda_{\rm max}$. Red line:
$\lambda_{\rm max, crit}$ where $f_{\rm acc} = 0.2$. Blue line:  $\lambda_{\rm
max, crit}$ where $f_{\rm acc} = 0.5$.  \label{fig:b1:compressive}}
\end{figure*}

\subsection{Hierarchical and localized   star formation}
To link the density structures of the gas to the spatial organization of the newly-formed stars, using maps of $\lambda_{\rm min}$, we identify coherent regions where $\lambda_{\rm min} < 0$, and study their properties. For each region,  we derive  a
mass $m_{\rm core}$ and a size $r_{\rm core}$. We removed
regions whose masses are smaller than the Jeans mass, as they would not be able
to collapse under self-gravity. Among the massive, super-Jeans regions, most have sizes
that we can barely resolve $l \lesssim 0.1 \;\rm pc$, and they contain a
wide range of masses ($0.3 - 30\;\rm M_{\odot}$)  (Fig. \ref{fig:b1:extensive}).
\
We call these regions ``islands under compressive tides'', which reflects their compactness in space. We find that 
a clump typically consists of a few tens of these ``islands''.  We use the following equation to estimate the number of stars formed in one of these islands 
\begin{equation}
  N_{*\;\rm island} =  \frac{N_{*, \rm total}}
  {N_{\rm island}} = \frac{M_{\rm gas} \;\eta_{\rm SF}\; m_{*, \rm mean}^{-1}}{N_{\rm island}} \approx 20 \;,  
\end{equation}   
where $N_{*, \rm total} = M_{\rm gas} \eta_{\rm SF}  m_{*, \rm mean}^{-1}$ 
is the total number of stars, $M_{\rm gas} \approx \; 1000\; M_{\odot}$ is the mass of the clump, $\eta_{\rm SF} = M_{\rm stars} /  M_{\rm gas}$ is the star formation efficiency of a cluster,  and $m_{*, \rm mean} = 0.3\; M_{\odot}$ 
 is the IMF-averaged stellar mass \citep{2001MNRAS.322..231K}. $N_{\rm island}$ is the number of islands. On average, around 20 stars will be formed in one island. These are called stellar groups.

 This spatial arrangement appears to be consistent with observational constraints of the initial condition of star formation. It has been well-established that binaries
born in massive star clusters has shorter periods.  
By modeling the orbital parameter distribution of binaries, one can infer the density of the environment where stars are born. Through this approach,  studies
\citep{2012A&A...543A...8M} finds that stars in clusters form in dense
regions of small sizes. A 1000 $M_{\odot}$ clumps such as the B1 region should from  a 250 $M_{\odot}$
cluster. According to previous estimates \citep{2012A&A...543A...8M}, the stars should be born in regions of $3000\;
M_\odot\;\rm pc^{-3}$ ($n_{\rm H_2} = 5 \times 10^5 \;\rm cm^{-3}$). We can relate this high stellar density to the high gas 
densities of these ``islands'' found in our observations. From our maps, most of
these islands are small ($l \lesssim 0.1\rm \;pc$), barely resolved, and contain
different masses. A typical, 0.1 pc-sized ``island'' of $10\; M_{\odot}$ has a
gas density of 2500 $M_{\odot}\;\rm pc^{-3}$, and this is very similar to the
stellar density inferred from observations. Although what we measure is the gas
density, and the corresponding stellar density can be lower,
 our ``islands'' are still growing in mass. The compactness of these ``islands'' can be related to the high stellar
density of clusters stars. In our case, the extensive tides confine star formation to these small, dense regions. 

The existence of these islands agrees with predictions of simulations \citep{2017MNRAS.467.1313V} carried out under the of the Global Hierarchical Collapse 
scenario \citep{2009ApJ...707.1023V}. To perform these simulations, the authors used supersonic turbulence to create a hierarchy of density structures,  which collapse to produce groups of stars. They related the stellar groups to the hierarchical density structure produced by turbulence. This spatial arrangement is very similar to what was seen from our maps, and the high densities of the gas clumps where groups form are in agreement with the high stellar density of stellar nurseries inferred from observations. 
 In addition, we find that gravity from the dense structures can suppress the ambient gas of lower densities from fragmenting, which limits the range of scales that can collapse, confining star formation to localized regions. We propose that it is this \emph{tidal-driven confinement} plays a critical role in sustaining the high-density condition where stars form.






\subsection{Accretion towards  dense regions}
Another consequence of this spatial arrangement is that to 
explain the observed star formation efficiency, transport of gas from 
regions under extensive tides to the dense ``islands'' is necessary. We use the following equation to describe the transport of gas from the clumps to the stars:
 \begin{equation}
   \frac{m_*}{m_{\rm gas}} = (1 - f_{\rm outflow}) f_{\rm infall} \;,
 \end{equation}
where $m_*$ is the stellar mass, $m_{\rm gas}$ is the gas mass, $(1 - f_{\rm
outflow})$ represents the effect of outflow  which can transport the infalling gas outwards \citep{2000ApJ...545..364M}, and $f_{\rm infall}$ is the
fraction of gas clump-scale gas that can contribute to the infall. Adopting  values consistent with literature results
(${m_*}/{m_{\rm gas}} = 0.25$ \citep{2016A&A...586A..68P}, $f_{\rm outflow} \approx 0.5$ \citep{2000ApJ...545..364M}), we require $f_{\rm
infall} \approx 50 \%$. To explain the current conversion efficiency, roughly half of the gas in a clump is expected to be to be
accreted. Since only  25\% of the gas in found in the islands under compressive tides, the remaining 25 \% likely contributed by the transport of gas from diffusion regions towards these islands.

Using our maps, we  can find out which part of the gas will be accreted. We note that the spatial distribution of $\lambda_{\rm max}$ has a
stratified structure, where gas located at the cluster centers have larger
$\lambda_{\rm max}$, which implies shorter evolution times ($t_{\rm acc} \approx
\lambda_{\rm max}^{1/2}$). This corresponds to the picture where gas located at cluster centers have shorter evolution times and will be accreted first. 
We can relate $f_{\rm infall}$ to $\lambda_{\rm max}$ by solving
\begin{equation}
  m_{\rm clump}  f_{\rm infall} = \int_{\lambda_{\rm max, crit}}^{\inf}\frac{ {\rm d} m(\lambda_{\rm max})}{{\rm d}{\lambda_{\rm max}}} {\rm d} \lambda_{\rm max}\;, 
\end{equation}
where $\frac{ {\rm d} m(\lambda_{\rm max})}{{\rm d}{\lambda_{\rm max}}}$ is the
mass-weighted distribution of $\lambda_{\rm max}$. After this, the locations of
gas with $\lambda_{\rm max} > \lambda_{\rm max, crit}$ can be mapped. This approach gives insight on which part of the gas is more likely to end up in stars. In reality, turbulence can leads to some additional mixing which is not accounted for. 
The
results at $\lambda_{\rm max, crit} = [0.2, 0.5]$ are plotted (Fig \ref{fig:b1:compressive}). The very
first part ($f_{\rm infall} = 0.2$) of the gas to be converted into stars are those
located in the cluster center under compressive tides, followed by gas with $0.2<f_{\rm infall}<0.5$, which
contains both gas at  the central regions and gas contained in filamentary
structures. To explain the observe star formation efficiency, accretion of gas located where tides are extensive is necessary at the late stage of the evolution.

\section{Conclusions}
One prevailing paradigm about star formation is that the collapse is
driven by gravity, balanced by processes such as turbulence and magnetic
fields \citep{2020SSRv..216...64K,2019ARA&A..57..227K}.  This scenario recognizes the role of gravity as the driver of the cloud evolution, but is over-simplified as the various ways gravity can act on the
gas are not differentiated. The long-range nature of gravity is also overlooked.

By analyzing the structure of the gravitational field using the
tidal tensor, we reveal the multiple ways through which gravity can act on the
gas. In dense regions, gravity derives collapse, and in their
surroundings, gravity suppresses fragmentation through extensive tides. We find that the vast majority of the gas study is under the influence of the extensive tides --  a crucial fact that has been overlooked so far. 
In addition, our analyses indicate that transport of gas towards the dense regions is necessary.

Our results provide crucial insights into the collapse process. In the turbulent
 fragmentation theory,
 \citep{2002ApJ...576..870P,2008ApJ...684..395H,2012MNRAS.423.2037H}, star
 formation arises from the collapse of the density fluctuations created by
 turbulence. In the competition accretion scenario \citep{2001MNRAS.323..785B},
 stars in a cluster share a  gas reservoir, and they acquire mass by accreting
 the gas competitively.  In our view, turbulence can be responsible for some
 initial density fluctuations, but gravity quickly takes over,  dictating the
 interactions between the high-density regions and their surroundings through
 tides. Accretion onto protostars can occur, but the competition is necessarily limited to
 those staying on the same islands. This picture where star formation occurs in
 a few more discrete ``islands'' is consistent with the picture from the Global
 Hierarchical Collapse scenario \citep{2009ApJ...707.1023V}.  We further
 discovered the compactness of these islands is the consequence of the mechanism of \emph{tidal-induced confinement}, and their high densities are maintained by accretion. \citep{2001MNRAS.323..785B}.  The picture of stellar groups  accrete gas as
  whole entities is reminiscent of the ``collaborative accretion''  scenario
 \citep{2014ApJ...782L...1E}.

Although tides were known since the time of Newton \citep{newton}, this is the first time where its dominant importance in controlling cloud evolution is revealed in detail. 
We expect the link between the density structure of the gas the role of gravity to be strengthened by future studies through which a clear view of the star formation process be established. 

\section*{Acknowledgements}
We thank the referee for careful readings of the paper and the comments, which greatly improved the structure of the paper. 
GXL acknowledges supports from NSFC grant K204101220130,
W820301904 and 12033005. 
\bibliographystyle{mnras}
\bibliography{paper.bib}
\appendix
\section*{Data availability}
Observational data: available in the link provided in \citet{2016A&A...587A.106Z}.
Simulation data (for testing): available at \url{https://starformmapper.org/project-index/}.

\section{Constructing 3D density distribution} \label{sec:reconstruction}
We use the method developed in our previous papers to reconstruct a 3D density
distribution. First, we decompose an input surface density map $I(x,y)$ into component
maps that contain structures of different scales $I_{l_i}(x,y)$, where $l_i$ is a range of scales, e.g. $2, 4, 8 ... 2**n$ pixels \citep{2022ApJS..259...59L}. In the second step, each component map $I_{l_i}(x,y)$ is converted into a cube $I_{l_i}(x,y, z)$, where a thickness $d = l_i$ is assigned. Finally, we combine these cubes to obtain a 3D density map. The whole procedure was described in \citet{2022MNRAS.514L..16L}. In our reconstructed density distributions, the density maximums
along different line-of-sights stay on the same plane.

\section{Gravitational potential and tidal tensor}\label{sec:phi}
To compute the gravitational potential and the tidal tensor, we first projected 
the data onto a rectangular grid. The gravitational potential is computed by solving the Poisson Equation
$\nabla^2 \phi = 4 \pi G \rho$. The computation is carried out in the Fourier
space, where we first obtain $\rho_k = \hat{f}(\rho(x, y, z))$. The gravitational potential in the Fourier space is computed via  $ \phi_k = 4 \pi G \rho_k k^{-2}$. The gravitational potential in the real space is computed through $\phi_(x, y, z) = \hat{f}^-1(\phi_k)$, where $\hat{f}$ is the Fourier transform and $\hat{f}^{-1}$ is the inverse Fourier transform. 

The tidal tensor, defined as
$T_{ij} = \partial^2 \phi / \partial i \partial j$, is computed at every location, and the eigenvalues of the tidal tensor as well as the eigenvector are computed using \texttt{scipy.linalg.eig} function provided in the \texttt{Scipy} package \citep{2020NatMe..17..261V}. The tidal tensor have three  eigenvalues, $\lambda_{\rm min}$, $\lambda_{\rm med}$ and $\lambda_{\rm max}$,
where  $\lambda_{\rm min} < \lambda_{\rm med} < \lambda_{\rm max}$. Note that
${\rm tr} (T_{ij}) = \lambda_{\rm min} + \lambda_{\rm med} + \lambda_{\rm max} =
4 \pi G \rho$.

When  computing the gravitational potential, we did not take the masses of the
YSOs (Young Stellar Objects) into account, as their contributions are negligible
in most cases. In the Perseus molecular cloud, the only region which contains a
large number of YSOs is the NGC1333 region. It contains a total of 1000
$M_{\odot}$ mass,  and around $100$ YSOs \citep{2015AJ....150...40Y}.
Assuming an IMF-averaged stellar mass of 3 $M_{\odot}$, the amount of mass of
stars is around 30 $M_{\odot}$, which is still small. 

\section{Validating using numerical simulations}\label{sec:comparison}
\begin{figure*}
  \includegraphics*[width = 1 \textwidth]{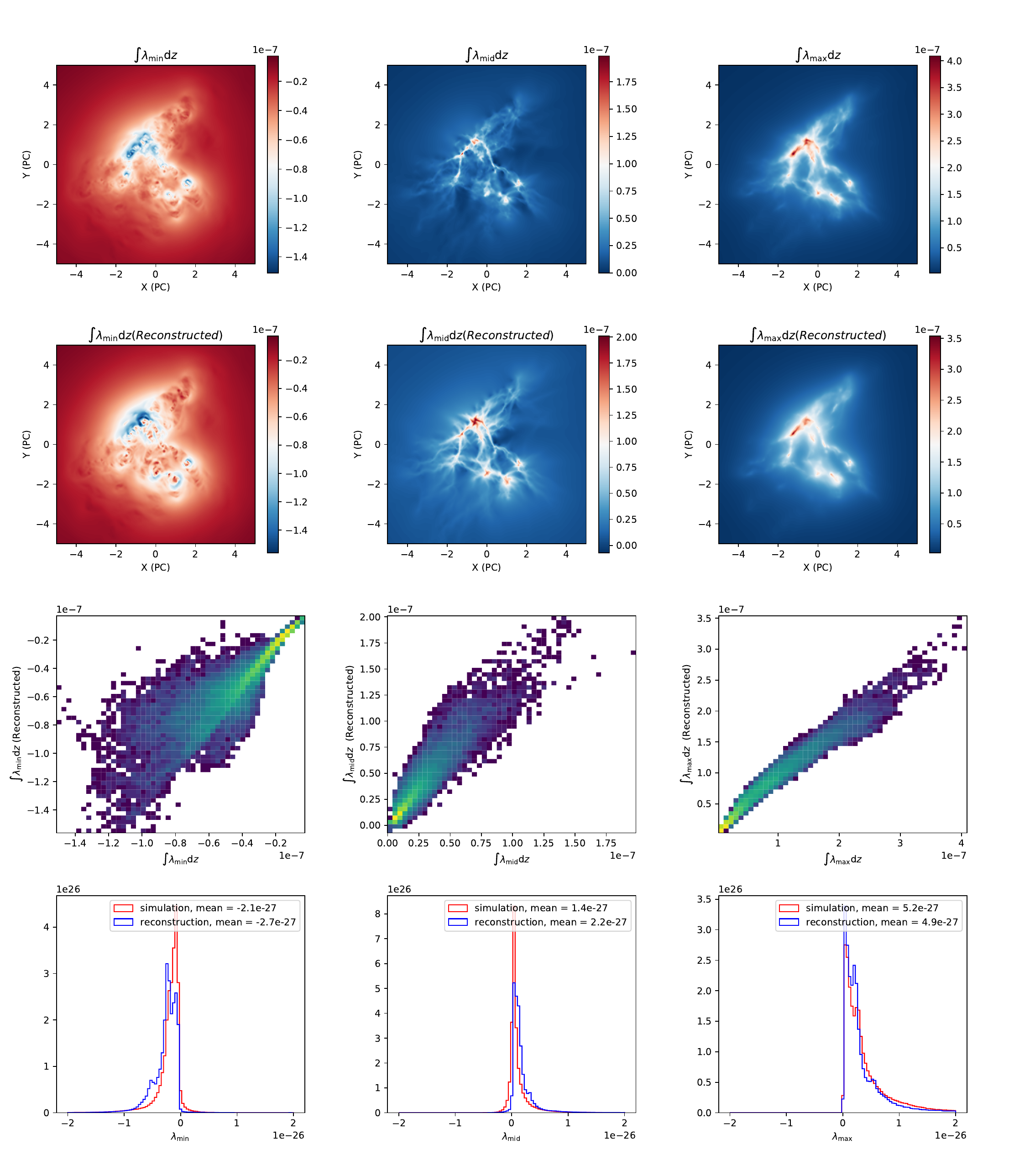}
  \caption{{\bf Comparisons of maps of eigenvalues of the tidal tensor.} In the first row, we plot integrated maps of the eigenvalues of the tidal tensor computed using the gas density map of a simulated cloud, where $\lambda_{\rm mean}(x, y) =  \int \lambda(x, y, z) {\rm d}z $. In the second row, we plot similar maps towards a 3D cloud reconstructed from a 2D surface density map. In the third row, we compare the original map values against the map values from the reconstructed density distributions. In the last row, we compare the mass-weighted distributions of the eigenvalues of the tidal tensor, where $<\lambda > = \int \lambda \rho  {\rm d}^3 x / \int \rho {\rm d}^3 x$.  
  \label{fig:comparison}}
\end{figure*}
The reconstructed density reconstructed resembles but is not identical to, the
original one. The validity of our results relies on the fact the results we obtained from 
the reconstructed cloud resemble those obtained from the real cloud. 

To verify our method, we use results from \citep{2019MNRAS.486.4622C} which
simulated the collapse of a cloud using the AREPO code
\citep{2010MNRAS.401..791S}, where they simulated the formation of a 1300 $M_\odot$ cloud with self-consistent hydrogen, carbon, and oxygen chemistry.

We compute the eigenvalues of the tidal tensor using
the original 3D density distribution, as well as the density structure from our
distribution. In Fig. \ref{fig:comparison}, we compare the spatial structures
and distributions of the eigenvalues of the tidal tensor. In the original data,
$<\lambda_{\rm min}> = -2.1 \times 10^{-27}$, $<\lambda_{\rm max}> =  5.2 \times
10^{-27}$. In our constructed data  $<\lambda_{\rm min}> = -2.7 \times
10^{-27}$, $<\lambda_{\rm max}> =  4.9 \times 10^{-27}$. $\lambda_{\rm min}$
differs by 25\% and $\lambda_{\rm max}$ differs by 8 \% \footnote{Here, the means are weighted by density, e.g.  the
$<\lambda > = \int \lambda \rho  {\rm d}^3 x / \int \rho {\rm d}^3 x$.}. In the original  data,
18\% of the gas stays under compressive tides, and in our reconstructed data
12\% of the gas stays under compressive tide. The mass fractions we derived are
accuracy to around 10 \%.

\bsp	
\label{lastpage}
\end{document}